# ASTROD AND ASTROD I – OVERVIEW AND PROGRESS

WEI-TOU NI

*Center for Gravitation and Cosmology, Purple Mountain Observatory,
Chinese Academy of Sciences, Nanjing, China, 21008, and
National Astronomical Observatories, Chinese Academy of Sciences, Beijing, China, 100012*
wtni@pmo.ac.cn

In this paper, we present an overview of ASTROD (Astrodynamical Space Test of Relativity using Optical Devices) and ASTROD I mission concepts and studies. The missions employ deep-space laser ranging using drag-free spacecraft to map the gravitational field in the solar-system. The solar-system gravitational field is determined by three factors: the dynamic distribution of matter in the solar system; the dynamic distribution of matter outside the solar system (galactic, cosmological, etc.) and gravitational waves propagating through the solar system. Different relativistic theories of gravity make different predictions of the solar-system gravitational field. Hence, precise measurements of the solar-system gravitational field test these relativistic theories, in addition to enabling gravitational wave observations, determination of the matter distribution in the solar-system and determination of the observable (testable) influence of our galaxy and cosmos. The tests and observations include: (i) a precise determination of the relativistic parameters β and γ with 3-5 orders of magnitude improvement over previous measurements; (ii) a 1-2 order of magnitude improvement in the measurement of $\dot{G}$; (iii) a precise determination of any anomalous, constant acceleration $A_a$ directed towards the Sun; (iv) a measurement of solar angular momentum via the Lense-Thirring effect; (v) the detection of solar g-mode oscillations via their changing gravity field, thus, providing a new eye to see inside the Sun; (vi) precise determination of the planetary orbit elements and masses; (viii) better determination of the orbits and masses of major asteroids; (ix) detection and observation of gravitational waves from massive black holes and galactic binary stars in the frequency range 50 μHz to 5 mHz; and (x) exploring background gravitational-waves. The baseline scheme of ASTROD is to have two spacecraft in separate solar orbits and one spacecraft near the Earth-Sun L1/L2 point carrying a payload of a proof mass, two telescopes, two 1-2 W lasers with spares, a clock and a drag-free system ranging coherently among one another using lasers. ASTROD I is a first step towards ASTROD. Its scheme is to have one spacecraft in a Venus-gravity-assisted solar orbit, ranging optically with ground stations with less ambitious, but still significant scientific goals.

## 1. Introduction

The distance determination of satellite laser ranging with two colors (two wavelengths) reaches millimeter accuracy. With the newer generation of lunar laser ranging,[1,2] the accuracy of distance determination will also reach millimeter accuracy. For the present lunar laser ranging accuracy of 2 cm, the result of testing relativistic gravity is comparable to that of interplanetary radio ranging. A comparison of these two methods for the solar-system tests of relativistic gravity is summarized in Table 1. A discussion of these solar-system tests is given in Ref. 3.

Table 1. Relativity-parameter determination from interplanetary radio ranging and from lunar laser ranging.

| Parameter | Meaning | Value from Solar System Determinations | Value from Lunar Laser Ranging |
|---|---|---|---|
| $\beta$ | PPN [4] Nonlinear Gravity | 1.000±0.003 [5] (perihelion shift with $J_2$ (Sun)=$10^{-7}$ assumed)<br>0.9990±0.0012 [6] (Solar-System Tests with $J_2$ (Sun)=(2.3±5.2)×$10^{-7}$ fitted)<br>1.0000±0.0001 [7] (EPM2004 fitting) | 1.003±0.005 [8]<br><br>1.00012±0.0011 [9, 10] |
| $\gamma$ | PPN Space Curvature | 1.000±0.002 [5] (Viking ranging time delay)<br>0.9985±0.0021 [6] (Solar-System Tests)<br>1.000021±0.000023 [10](Cassini S/C Ranging)<br>0.9999±0.0001 [7] (EPM2004 fitting) | 1.000±0.005 [8] |
| $K_{gp}$ | Geodetic Precession | | 0.997±0.007 [8]<br>0.9981±0.0064 [9] |
| E | Strong Equivalence Principle | | (3.2±4.6)×$10^{-13}$ [8]<br>(-2.0±2.0)×$10^{-13}$ [9, 11] |
| $\dot{G}/G$ | Temporal Change in G | (2±4)×$10^{-12}$/yr [12] (Viking Lander Ranging)<br>±10×$10^{-12}$/yr [13] (Viking Lander Ranging)<br>±2.0×$10^{-12}$/yr [14](Mercury & Venus Ranging)<br>±(1.1-1.8)×$10^{-12}$/yr [15] (Solar-System Tests) | (1±8) × $10^{-12}$/yr [8]<br>(0.4±0.9)× $10^{-12}$/yr [9] |

In broad terms, both solar-system radio ranging and lunar laser ranging have tested relativistic-gravity effects to $10^{-3}$-$10^{-4}$. When laser ranging is extended to the whole solar system, the precision of relativistic gravity tests will be dramatically improved. After many years of research and development of lunar laser ranging and satellite laser ranging, the laser ranging technology is ripe for deep-space laser ranging.

Recently, interplanetary laser ranging was demonstrated by MESSENGER (MErcury Surface, Space ENvironment, GEochemistry, and Ranging).[16-18] The MESSENGER spacecraft, launched on 3 August 2004, is carrying the Mercury Laser Altimeter (MLA) as part of its instrument suite on its 6.6-year voyage to Mercury. Between 24 May, 2005 and 31 May, 2005 in an experiment performed at about 24 million km before an Earth flyby, the MLA on board MESSENGER spacecraft performed a raster scan of Earth by firing its Q-switched Nd:YAG laser at an 8 Hz rate. Pulses were successfully received by the 1.2 m telescope aimed at the MESSENGER spacecraft in the NASA Goddard Geophysical and Astronomical Observatory at Gaddard Space Flight Center (GSFC) when the MLA raster scan passed over the Earth station. Simultaneously, a ground based Q-switched Nd:YAG laser at GSFC's 1.2 m telescope was aimed at the MESSENGER spacecraft. Pulses were successfully exchanged between the two terminals. From this two-way laser link, the range as a function of time at the spacecraft over 2.39 × $10^{10}$ m (~ 0.16 AU) was determined to ± 0.2 m (± 670 ps): a fractional accuracy of better than 1 × $10^{-11}$.

A similar experiment was conducted by the same team to the Mars Orbiter Laser Altimeter (MOLA) on board the Mars Global Surveyor (MGS) spacecraft in orbit about Mars.[18] At that time, the MOLA laser was no longer operable after a successful topographic mapping mission at Mars. The experiment was one way (uplink) and the MOLA detector saw hundreds of pulses from 8.4 W Q-switched Nd:YAG laser at GSFC.

These interplanetary laser ranging demonstrations are encouraging. Compared with MESSENGER's 2-way ranging, 10 ps ranging accuracy and using distances which are 10 times longer, as aimed in ASTROD I, will achieve a fractional accuracy of $10^{-14}$.

Here, we give an overview of the ASTROD (Astrodynamical Space Test of Relativity using Optical Devices) and ASTROD I mission concepts and progress in their study.

## 2. ASTROD

The general concept of ASTROD (Astrodynamical Space Test of Relativity using Optical Devices) is to have a constellation of drag-free spacecraft navigate through the solar system and range with one another using optical devices to map the solar-system gravitational field, to measure related solar-system parameters, to test relativistic gravity, to observe solar g-mode oscillations, and to detect gravitational waves.

A baseline implementation of ASTROD is to have two spacecraft in separate solar orbits, each carrying a payload of a proof mass, two telescopes, two 1–2 W lasers, a clock and a drag-free system, together with a similar spacecraft near Earth at one of the Lagrange points L1/L2.[19-21] The three spacecraft range coherently with one another using lasers to map solar-system gravity, to test relativistic gravity, to observe solar g-mode oscillations, and to detect gravitational waves. Distances between spacecraft depend critically on solar-system gravity (including gravity induced by solar oscillations), underlying gravitational theory and incoming gravitational waves. A precise measurement of these distances as a function of time will enable the cause of a variation to be determined. After 2.5 years, the inner spacecraft completes 3 rounds about the Sun, the outer spacecraft 2 rounds, and the L1/L2 spacecraft (Earth) 2.5 rounds. At this stage two spacecraft will be on the other side of the Sun, as viewed from the Earth, in the optimum configuration for the Shapiro time delay experiment. The spacecraft configuration after 700 days from launch is shown in Fig. 1. Whenever there is no ambiguity, we denote this baseline implementation as ASTROD also.

The accuracy and the precision of the measurement depend on the timing accuracy, the inertial sensor/accelerometer noise and the laser/clock stability. As for the timing accuracy, an event timer has been designed at OCA (Observatoire de la Côte d'Azur) in the framework of both the T2L2 (Time Transfer by Laser Link) project and the laser ranging activities.[22,23] At present, the prototype of the OCA timer is fully operational and has a precision of better than 3 ps, a linearity error of 1 ps rms and a time stability of less than 0.01 ps over 1000 s with dead time of less than 10 μs. For a mission within the next 10-20 years, a timing accuracy of better than 1 ps (300 μm in terms of range) is anticipated. In coherent interferometric ranging, timing events need to be generated by a modulation/encoding technique or by superposing timing pulses on the CW laser light. The interference fringes serve as consecutive time marks. With timing events aggregated to a normal point using an orbit model, the precision can reach 30 μm in range. The effective range precision for parameter determination could be better, reaching 3-10 μm using orbit models. Since the ASTROD range is typically of the order of 1-2 AU (1.5-3 × $10^{11}$ m), a range precision of 3 μm will give a fractional precision of distance determination of $10^{-17}$. Therefore, the desired clock accuracy/stability is $10^{-17}$ over 1000 s travel time. Optical clocks with this accuracy/stability are under development, being a hot discussion topic in the First ESA Optical Clock Workshop that was held in June, 2005. A space optical clock is under development for the Galileo project. This development would pave the road for ASTROD to use optical clocks.[24-26] The present range precision for radio tracking is a couple of meters. The improvement of ASTROD will be 5 orders of magnitude, allowing relativistic gravity test in the solar system to be tested to ~1 ppb (part per billion).

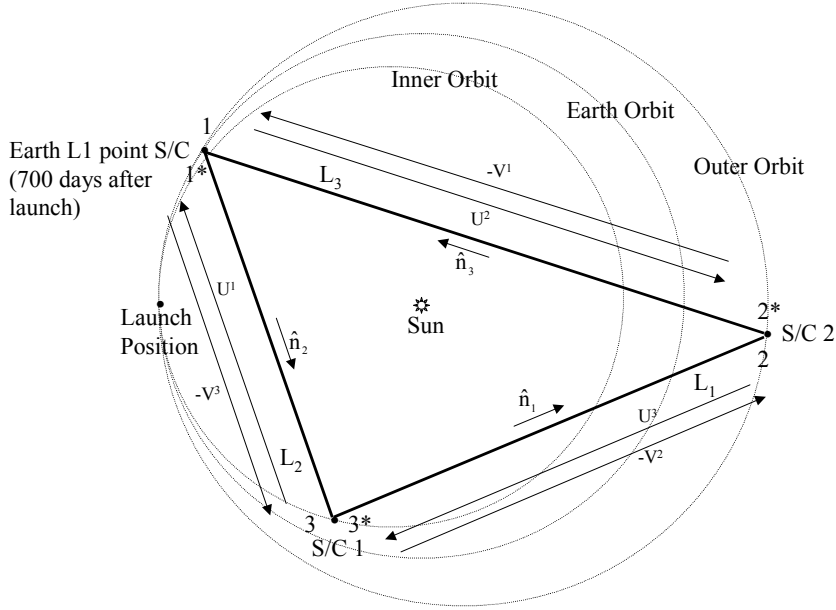

Fig. 1. A schematic ASTROD configuration (baseline ASTROD after 700 days from launch)

Having reached the 1 ppb realm in range measurement, high precision drag-free performance is a must to guarantee that the test masses of the ASTROD spacecraft follow geodesics as closely as possible in the solar system, to extract meaningful fundamental physics. In terms of drag-free performance, ASTROD aims for residual test mass acceleration noise of

$$S_{\Delta a}^{1/2}(f) = (0.3\text{-}1) \times 10^{-15} \, [1 + 10 \times (f/\, 3 \text{ mHz})^2] \text{ m s}^{-2} \text{ Hz}^{-1/2}, \qquad (1)$$

in the frequency range 0.1 mHz $< f <$ 100 mHz. The ASTROD (Eq. (1) with the more stringent choice) noise requirement curve is compared to the noise requirement curves of LISA,[27] LISA Pathfinder LISA Technology Package (LTP),[28] and ASTROD I (see next section) in Fig. 2. The ASTROD requirement is more stringent than that of LISA at 0.1 mHz by a factor of 3-10. This improvement must be achieved by using capacitive sensing with larger gaps or by going to optical methods.[29] The concept of a laser metrology inertial sensor was reported to the Fundamental Physics in Space Symposium in 1995 in London[30] and the COSPAR General Assembly in 1996 in Birmingham.[31,32] A reference-proof mass configuration for ASTROD is shown in Fig. 3. Dummy masses are included to minimize the gravitational gradient at the position of the proof mass. Recently, laboratory developments for an interferometric sensor for spacecraft drag-free control,[33] and an advanced optical gravitational reference for high precision space interferometers[34] have been reported. All these experiments need to use lasers as standard rods. This means that they need to be calibrated using frequency standards. The developments of optical clocks and optical combs will make things simpler in these space missions. A detailed discussion of current prospects for the ASTROD inertial sensor is presented in reference [35].

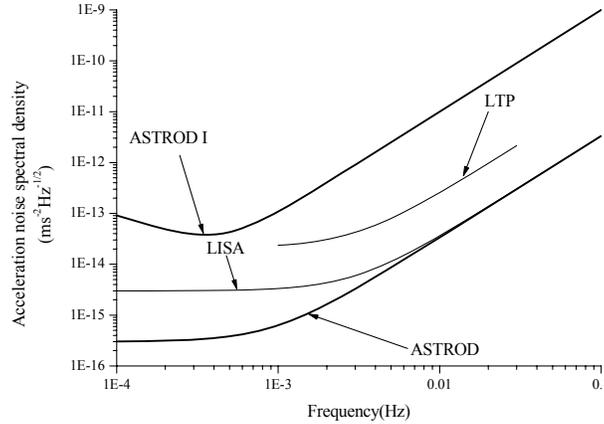

Fig. 2. A comparison of the target acceleration noise curves of ASTROD, LISA, the LTP and ASTROD I.

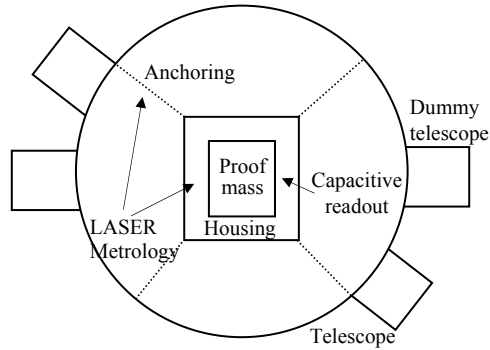

Fig. 3. A reference-proof mass configuration with capacitive readout.

In our preliminary orbit simulation, with the required accelerometer noise given by Eq. (1) and with 1 ps timing error, we obtain that the uncertainty in measuring $\gamma$ is in the $(1-3) \times 10^{-9}$ range, and that for $J_2$ is in the $10^{-10}$ range. In 2000, we reported an orbit design for ASTROD with both inner and outer spacecraft apparent positions near the solar superior conjunction as seen from the Earth or L1/L2 spacecraft between 800 to 1050 days and between 2600 and 2850 days after launch in 2005.[36] During these days, the Shapiro time delay (and $\gamma$) can be measured precisely and the correlation of the relativistic parameters $\beta$ and $\gamma$ decreased so that the uncertainty in $\beta$ can be reduced. In 2002 we reported a similar orbit design for 2015 launch.[37,20] Here, we report the simulation for a closely related 2015 launch orbit design[38,39] similar to the one reported in ref.'s [37, 20] with (i) timing error 1 ps and (ii) drag-free acceleration noise $1 \times 10^{-15}$ m s$^{-2}$ Hz$^{-1/2}$ at 0.1 mHz [the more relaxed choice of equation (1)]. The uncertainties of γ, β, J$_2$ and Ġ/G as functions of epoch after launch for this orbit simulation are shown in Fig. 4.[38,39] As we can see from the diagram for γ, its uncertainty reaches 1.24 parts per billion (10$^{-9}$) after 1200 days. The uncertainties after 3000 days for the five parameters γ, β, J$_2$, Ġ/G and A$_a$ are listed in Table 2. In this simulation, we used 31 parameters in the fitting, i.e., ($x_0^{inner}$, $y_0^{inner}$, $z_0^{inner}$, $v_{x0}^{inner}$, $v_{y0}^{inner}$, $v_{z0}^{inner}$, $x_0^{outer}$, $y_0^{outer}$, $z_0^{outer}$, $v_{x0}^{outer}$, $v_{y0}^{outer}$, $v_{z0}^{outer}$, M$_{Sun}$, M$_{Mercury}$, M$_{Venus}$, M$_{Earth}$, M$_{Moon}$, M$_{Mars}$, M$_{Jupiter}$, M$_{Saturn}$, M$_{Uranus}$, M$_{Neptune}$, M$_{Pluto}$, M$_{Ceres}$, M$_{Pallas}$, M$_{Vesta}$, γ, β, J$_2$, Ġ/G, A$_a$). Six parameters are for the initial position and velocity of the inner spacecraft; six for those of the outer spacecraft; eleven for the masses of Sun, Mercury, Venus, Earth, Moon, Mars, Jupiter, Saturn, Uranus, Neptune and Pluto; three for the masses of Ceres, Pallas and Vesta; two for the relativistic parameters γ and β; one for the solar quadrupole parameter J$_2$; one for Ġ/G; and one for the anomalous Pioneer acceleration A$_a$. In calculating the partials, we linearize the equations of motion around the fiducial trajectory with deviations small enough so that we can neglect the nonlinear errors but large enough so that numerical round-off errors are not important. In the fitting, we use the Kalman sequential filtering method and the 2 parameter post-Newtonian ephemeris framework. A full simulation needs to use the post-post-Newtonian ephemeris framework and to include post-post-Newtonian parameters. We are working towards this and developing this framework.[40-42] However, the basic results of this simulation are correct.

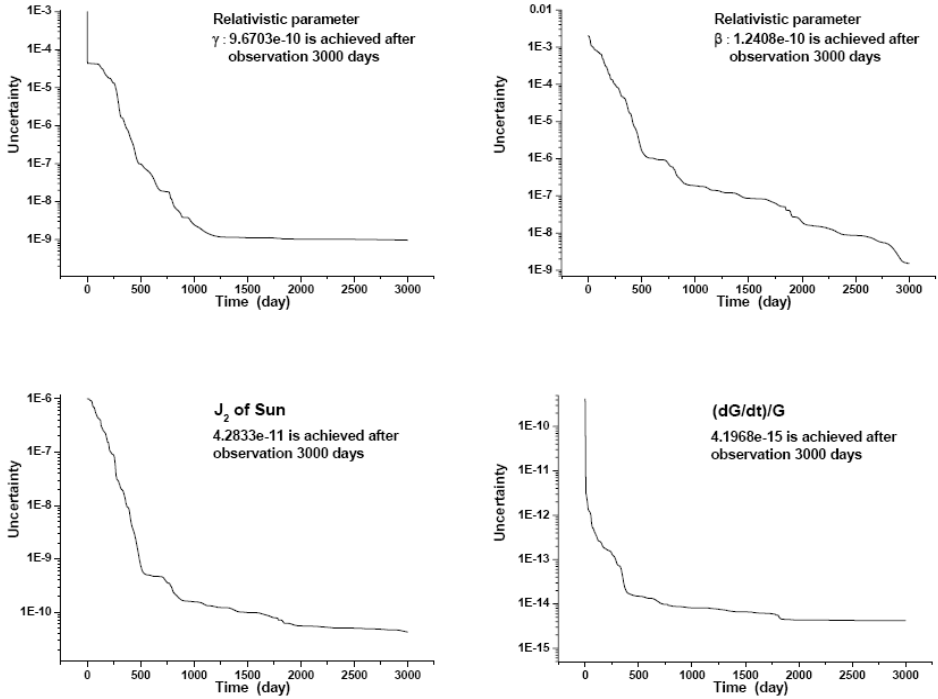

Fig. 4. Uncertainties of γ, β, J$_2$ and Ġ/G as functions of epoch for a 2015 launch orbit choice.[38,39] The unit of ordinate in the Ġ/G diagram is yr$^{-1}$.

Table 2. Uncertainties for 5 parameters of a 2015 ASTROD orbit simulation.[38,39]

| Parameter | Meaning | Initial value | Uncertainty after 3000 days |
|---|---|---|---|
| β | PPN Nonlinear Gravity | 1 | $1.24 \times 10^{-9}$ |
| γ | PPN Space Curvature | 1 | $9.67 \times 10^{-10}$ |
| $J_2$ | Quadrupole parameter of Sun | $2 \times 10^{-7}$ | $4.28 \times 10^{-11}$ |
| $\dot{G}/G$ [$yr^{-1}$] | Temporal Change in G | 0 | $4.20 \times 10^{-15}$ |
| $A_a$ [m s$^{-2}$] | Anomalous Poineer acceleration | 0 | $1.72 \times 10^{-17}$ |

The algorithms for unequal-arm noise cancellation of Armstrong, Estabrook and Tinto[43] are applicable to ASTROD (Fig. 1). The gravitational-wave sensitivity of ASTROD is shifted toward lower frequencies compared with that of LISA because of ASTROD's longer armlengths.[19]

Both LISA and ASTROD use interferometric laser ranging, and the Doppler effects on transmitted and received frequencies need to be addressed. LISA's strategy is to minimize armlength variations and relative velocity of the spacecraft. Orbit optimization methods and schemes for LISA are presented in references [44] and [45]. For ASTROD, the armlength changes of the 3 spacecraft are of the same order as the distances between the 3 spacecraft and the relative velocities go up to 70 km/s with line-of-sight velocities varying from -20 to 20 km/s. For 1064 nm (532 nm) laser light, the Doppler frequency change goes up to 40 (80) GHz. Hence for ASTROD, we must use a different strategy.[26] The recent development of optical clocks and frequency synthesizers using the optical comb[46] makes this heterodyne problem more tractable.

Space optical clocks and optical comb frequency synthesizer technologies are important elements in the realization of the ASTROD target sensitivity. Another use of the optical clock and optical comb frequency synthesizer is to calibrate the optical metrology for ASTROD. This is important for the laser metrology inertial sensor and for monitoring distances inside spacecraft, to correct local gravity changes due to, for example, thermal effects. All these measurements use lasers as standard rods. They need to be calibrated using optical frequency standards. The advent of optical clocks and optical combs in space has made the ASTROD mission feasible, simplifying the experimental design.

The development of ultra-high precision optical clocks is key to realization of ASTROD's scientific goals. We anticipate the development of space-borne ultra-high precision optical clocks by the optical community. The intrinsic accuracy of optical clocks will soon reach $10^{-17}$. It is reasonable to expect that optical clocks of this accuracy will be space qualified within the time-frame of the ESA Cosmic Vision (2015–2025), program.

The ASTROD spacecraft orbit around the Sun once every period and therefore there are orbital-period modulations both in amplitude and phase relative to the Sun from gravity strains induced by solar oscillations, in contrast to the strains induced from gravitational waves, coming from outside the solar system. Because of these orbital modulations, ASTROD will be able to separate the effects due to the gravity of solar g-mode oscillations from those due to the gravitational waves coming from outside the solar system. Among all experiments and proposals, ASTROD has the highest sensitivity to solar g-mode oscillations.[19,47] The old sensitivity goal of ASTROD to $\ell = 2$, $m = 2$ solar oscillations[20,47] touches the estimates of Kumar, Quartaert and Bahcall.[48,47] Now, because we have raised the requirement on acceleration noise by a factor of 3-10, the new sensitivity goal is 3-10 times more stringent. We will reach 3-10 times deeper into the estimates of Kumar, Quartaert and Bahcall. As this kind of mission has the best chance of detecting low $\ell$ solar oscillations and g-mode solar oscillations, a shared/dedicated mission is given serious consideration in the helioseismology community.[49] With this in mind, we are considering an orbit design of the ASTROD inner spacecraft similar to that of the ASTROD I spacecraft orbit which goes to about 0.5 AU in perihelion after the gravity assistance from 2 Venus flybys. The original orbit design for ASTROD inner spacecraft has a perihelion distance of 0.77 AU. The gravitational acceleration due to quadrupole modes is proportional to $r^{-4}$ where $r$ is the heliocentric distance of the spacecraft. The relative acceleration is of the same order as the individual accelerations for the spacecraft due to solar oscillations since the spacecraft are widely separated. Hence the gain in sensitivity is about $(0.77/0.5)^4 = 5.6$ fold; more accurate numbers will require more detailed calculations. With the new sensitivity goal, ASTROD should reach 20–60 times deeper into the estimates of Kumar, Quartaert and Bahcall and will have a good chance to observe g-mode and low-$\ell$ mode oscillations.

One of the goals of ASTROD is to measure the solar Lense-Thirring effect to determine the solar angular momentum. To measure the solar Lense-Thirring effect, we need the difference, $t_1-t_2$, in the two round trip propagations, L1/L2 S/C (Spacecraft) → S/C 1 (Spacecraft 1) → S/C 2 (Spacecraft 2) → L1/L2 S/C and L1/L2 S/C → S/C 2 → S/C 1 → L1/L2 S/C. Newtonian calculations of $t_1-t_2$ for 800–1034 days after launching gives $t_1-t_2$ about 10 ms. The Lense-Thirring effect variation for this period of time is about 100 ps and has a unique signature (compared to those of Shapiro time delay and other effects). For a laser stability of $10^{-15}-10^{-13}$, the accuracy for measuring the round-trip time difference $t_1-t_2$ (about 10 msec) is 10 as (attosecond) to 1 fs and the sensitivity of measurement of the Lense-Thirring effect is $10^{-7}-10^{-5}$ by means of the interferometric measurement. Based on a preliminary theoretical model, we expect that the Lense-Thirring effect can be determined to 0.1% or better.[50] Since the solar Lense-Thirring effect is proportional to the solar angular momentum, this measurement will give a value of the solar angular momentum with 0.1% accuracy or better. At present, there is about 5% uncertainty in the solar angular momentum modelling.

With the precision requirement of this mission, drag-free is a must. A Field Emission Electric Propulsion (FEEP) system in the μN range is under development in Centrospazio (Italy) and by the Austrian Research Centre Seibersdorf (ARC) for LISA.[27] The main issue for FEEPs is lifetime. ESA is dealing with this issue. More recently, development of Colloidal Micro-Newton Thrusters (CMNTs) was started by Busek Co. with support from JPL in testing and design.[51] The accelerometer/inertial sensor for LISA Pathfinder[28] is under development in the University of Trento. The LISA Pathfinder Technological Demonstration Mission of ESA is to be launched in 2009. For the accelerometer/inertial sensor design of ASTROD, an absolute laser metrology system is desired. This may push the noise down, in particular in the lower frequency region. Since ASTROD would be launched after LISA, ASTROD can inherit from the drag-free technologies developed for LISA: many of the developments made for the LISA[27] can be used in ASTROD, for example, the laser and telescope. Due to the absolute requirement for the drag-free system, the requirements for the ASTROD laser metrology system are demanding.[30,32] ASTROD needs monitoring the positions of various parts of the spacecraft, to facilitate gravitational modelling, and a proof mass system to be specially

designed.[29,31]

To reach a distance of more than 2 AU, coherent weak light phase locking to 100 fW incoming light is needed. For 100 fW ($\lambda$ = 1064 nm), there are still $5 \times 10^5$ photons/sec. This is enough for 100 kHz frequency tuning. In Tsing Hua University, 2 pW weak-light phase-locking with a 0.2 mW local oscillator has been achieved.[52] With pre-stabilization of lasers, improved balanced photodetection and lowered electronic circuit noise, the intensity goal should be readily achieved. The future should be focussed on offset phase locking, frequency-tracking and modulation-demodulation to mature the experimental technique for weak light. Weak-light phase-locking is also important for the deep space continuous-wave optical communication.

## 3. ASTROD I

ASTROD I is the first step towards the ASTROD mission. This mission concept has one spacecraft carrying a payload of a telescope, five lasers, and a clock together with ground stations (ODSN: Optical Deep Space Network) to test the optical scheme of interferometric and pulse ranging and still give important scientific results.[53,54]

The basic scheme of the ASTROD I space mission concept is to use two-way laser interferometric ranging and laser pulse ranging between the ASTROD I spacecraft in solar orbit and deep space laser stations on Earth to improve the precision of solar-system dynamics measurements, solar-system constants and ephemeris, to measure relativistic gravitational effects and to test the fundamental laws of spacetime more precisely, and to improve the measurement of the time rate of change of the gravitational constant.

A schematic of the payload configuration of ASTROD I is shown in Fig. 5.[53] The cylindrical spacecraft with diameter 2.5 m and height 2 m has its surface covered with solar panels. In orbit, the cylindrical axis is perpendicular to the orbital plane with the telescope pointing toward the ground laser station. The effective area to receive sunlight is about 5 m$^2$ and this can generate over 500 W of power. The total mass of spacecraft is 300-350 kg. That of the payload is 100–120 kg with a science data rate of 500 bps. A model using the actual dimensions and weight, for the main constituent parts of the ASTROD I spacecraft was developed for a simulation of the electrostatic charging of the ASTROD I test mass and is described in reference [55].

The spacecraft is 3-axis stabilized. It contains a 3-axis drag-free proof mass (PM) at the center and the spacecraft is forced to follow this proof mass using micro-thrusters. The drag-free performance requirement is below $10^{-13}$ m s$^{-2}$ Hz$^{-1/2}$ at 100 µHz to 1 mHz (3-axis) (Fig. 2). This performance is more than 10 times less stringent than the LISA drag-free system requirement. A $50 \times 50 \times 35$ mm$^3$ rectangular parallelepiped proof mass using Au-Pt alloy of low magnetic susceptibility ($< 5 \times 10^{-5}$) is planned. Titanium housing for the proof mass will remain at a vacuum pressure of less than 10 µPa. Capacitance sensing for the proof mass will be implemented for all six degrees of freedom. The titanium housing together with sensitive optical components is fixed on the optical bench. The optical bench is thermally stabilized. The laser ranging is between a fiducial point in the spacecraft and a fiducial point in the ground laser station. The fiducial point in spacecraft can be a separate entity which has a definite positional relation with respect to the proof mass. Incoming light will be collected using a 380-500 mm diameter f/1 Cassegrain telescope. This telescope will also transmit light from spacecraft, with $\lambda$/10 outgoing wavefront quality, to Earth. Ground laser stations will be similar to the present lunar laser ranging (LLR) stations or large satellite laser ranging (SLR) stations. At Yunnan Astronomical Observatory in Kunming, there is a large satellite laser ranging station with a 1.2 m azimuth-elevation reflection telescope. This station is being considered for use as a deep space laser station, to transmit and receive deep space laser signals.[56] Adaptive optics are under consideration for the telescope of the laser station. Weak-light phase locking has reached 2 pW.[52] The requirement on the strength of incoming light for phase locking on board the spacecraft is 100 fW. We are working toward this goal at present. A preliminary optics design for ASTROD I has been worked out in reference [57].

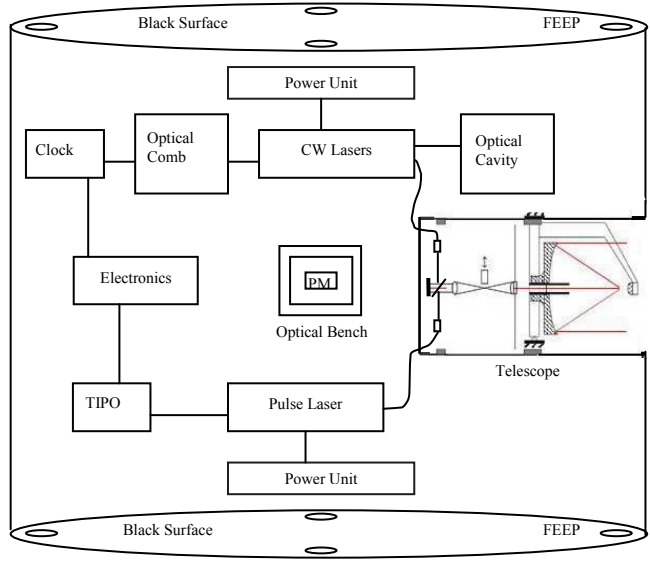

Fig. 5. A schematic diagram of payload configuration for ASTROD I.

The test (proof) mass in the center of Fig. 5 will be surrounded by electrodes on all six sides, to capacitively sense its motion relative to the spacecraft. Micro-thrusters on the spacecraft will then be used to force it to follow the test mass. The ASTROD I residual acceleration noise target is

$$S_{\Delta a}^{1/2}(f) = 3 \times 10^{-14} \, [(0.3 \text{ mHz} / f) + 30 \times (f / 3 \text{ mHz})^2] \text{ m s}^{-2} \text{ Hz}^{-1/2}, \qquad (2)$$

over the frequency range of 0.1 mHz < f < 100 mHz (Fig. 2). The strategy here is to have a moderate requirement compared with LISA, and still have important scientific goals in astrodynamics, solar-system measurement, and relativistic gravity.[53,58] The main parameter-values satisfying this requirement are listed in Table 2, compared with those of LISA.[58] The values for LISA are quoted from the current error estimates by Stebbins et al.,[59] apart from the thruster noise, taken from [27]. The requirement on the fluctuation of temperature in spacecraft (S/C) is much less stringent than that of LISA. Since the distance between ASTROD I and the Sun varies over the orbit and reaches 0.5 AU, this relaxation is important for realization of the thermal targets.

Table 2. Relaxed ASTROD I parameter values in comparison
with LISA at frequency of 0.1 mHz.[58]

| $f$ (frequency) = 0.1 mHz | ASTROD I | LISA |
|---|---|---|
| Maximum charge build-up: $q$ [C] | $10^{-12}$ | $10^{-13}$ |
| Magnetic susceptibility: $\chi_m$ | $5 \times 10^{-5}$ | $3 \times 10^{-6}$ |
| Magnetic remanent moment: $\|M_r\|$ [A m$^2$] | $1 \times 10^{-7}$ | $2 \times 10^{-8}$ |
| Residual gas pressure: $P$ [Pa] | $10^{-5}$ | $3 \times 10^{-6}$ |
| Electrostatic shielding factor: $\xi_e$ | 10 | 100 |
| Thruster noise [μN Hz$^{-1/2}$] | 10 | a few |
| Fluctuation of temperature in S/C: $\delta T_{S/C}$ [K Hz$^{-1/2}$] | 0.2 | 0.004 |
| Voltage difference between average voltage across opposite faces and voltage to ground: $V_{0g}$ [V] | 0.05 | 0.01 |
| Voltage difference between opposite faces: $V_d$ [V] | 0.01 | 0.005 |
| Fluctuation of voltage difference across opposite faces: $\delta V_d$ [V Hz$^{-1/2}$] | $10^{-4}$ | $10^{-5}$ |
| Residual dc voltage on electrodes: $V_0$ [V] | 0.1 | 0.01 |

A torsion pendulum study for a prototype inertial sensor for ASTROD I is reported in [60]. In this prototype torsion pendulum, the twist motion of the test mass is monitored and servo-controlled. The sensitivity of the electrostatic servo-controlled actuator is calibrated based on the elastic torque of the torsion fibre, and the torque resolution of the servo-controlled torsion pendulum comes to $2 \times 10^{-12}$ N m Hz$^{-1/2}$ at 3 mHz.

High-energy cosmic rays and solar energetic particles (SEP's) easily penetrate the light structure of spacecraft transferring heat, momentum and electrical charge to the test mass.[61] Electrical charging is the most significant of these disturbances. Any charge on a test mass will interact with the surrounding conducting surfaces through Coulomb forces. Further, motion of the charged test mass through magnetic fields will give rise to Lorentz forces. To limit the acceleration noise associated with these forces and meet the residual noise requirement, the test mass must be discharged in orbit. A charging simulation is reported in [55]. The charging of the ASTROD I test mass by cosmic ray protons and alpha particles ($^3$He and $^4$He) has been simulated using the GEANT4 toolkit. The spacecraft model used in [55] predicted a net charging rate of 33.3 e$^+$/s at solar minimum (12.7 e$^+$/s at solar maximum). We have also considered an additional net charging rate contribution of 1.4 e$^+$/s at solar minimum (0.9 e$^+$/s at solar maximum), due to particle species that were not included in this model, plus a potential additional ~28.4 e$^+$/s at solar minimum (17 e$^+$/s at solar maximum), due to kinetic low energy secondary electron emission. There is an additional uncertainty of ± 30% in the Monte Carlo net charging rate, due to uncertainties in the cosmic ray spectra, physics models and geometry implementation. The worst case charging rate is estimated to be 73.8 e$^+$/s.

The ASTROD I acceleration noise limit target is $10^{-13}$ m s$^{-2}$ Hz$^{-1/2}$ at 0.1 mHz, which is less stringent than the LISA requirement. The magnitudes of the Coulomb and Lorentz acceleration noise associated with test mass charging increase with decreasing frequency. At the lowest frequency in the ASTROD I bandwidth, 0.1 mHz, they are estimated at $2.8 \times 10^{-15}$ m s$^{-2}$ Hz$^{-1/2}$ and $2.8 \times 10^{-15}$ m s$^{-2}$ Hz$^{-1/2}$, respectively, both well below the acceleration noise target. However, variations in the test mass charging rate will alter the spectral description of the coherent Fourier components associated with charging. This is considered in reference [55] and the schemes of discharging of reference [62] can be applied to ensure that the spurious Fourier components do not compromise the quality of the science data of the ASTROD I mission.

The charging process of the ASTROD I test mass by SEPs (Solar Energetic Particles) has also been simulated using the GEANT4 toolkit, and the charging rate can be much larger than the values due to GCR (Galactic Cosmic Ray) proton flux at solar maximum and solar minimum.[63] However, the charge management hardware described in reference [62] could be used to discharge a test mass even during solar events, provided safe operation could be ensured. Better understanding of spacecraft environment will ensure better simulation, and a particle monitor on board spacecraft will be able to monitor environment and give useful information about interplanetary space.[64,65]

One aspect of our future work on this aspect of the project will be to implement a more complete list of physics processes in the GEANT4 simulation. We will also include the detailed study of other potential charging mechanisms, such as X-ray processes and photon emission. The effect of cosmic ray fluxes of particle species not included in this the present simulation needs to be verified for the ASTROD I geometry. Further, we will evaluate the variation in the ASTROD I test mass charging rate over the solar cycle, and its variation due to modulation of cosmic ray flux over the ASTROD I orbit. The ASTROD I orbit varies between about 0.5 AU to 1 AU, in radial distance from the Sun, and between $\pm 1°$, in latitude from the ecliptic plane.

In orbit design and orbit simulation for ASTROD I, we use the method used for ASTROD.[66,67] We note that for the Venus swing-by, to obtain the gravity assist to reach the other side of the Sun earlier, there is a launch window about every 584 days (synodic period of Venus).[68] There is a launch window in 2012, 2013 and 2015. In the following, we use the 2013 orbit as an example.[69] In the numerical integration used to find such an orbit, we take the following effects into account: (i) Newtonian and post-Newtonian point mass gravitational interaction between every two bodies including the Sun, nine planets, moon, three big asteroids (Ceres, Pallas and Vesta); (ii) Newtonian attraction between a body with gravitational multipoles including Sun (J2), Earth (J2, J3, J4) and Venus (J2), and one of the others as point masses; (iii) the spacecraft is treated as a test body in the numerical integration. The orbit we found started at 23:45:36 on October 18, 2012 (JD 2,456,584.490,000) and is shown in Fig. 6 in the X-Y plane of the heliocentric ecliptic coordinate system. Two Venus swing-bys are at 112.2 days and 336.6 days after launch. The times that the apparent position of spacecraft reaches the opposite side of the Sun are at 367.1 day and 697.6 day from launch.

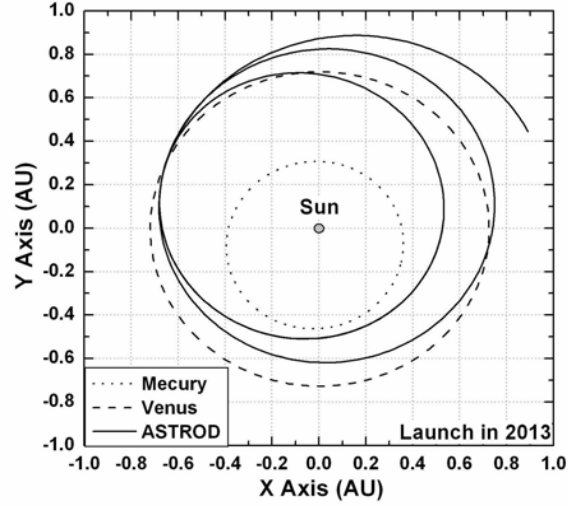

**Fig. 6.** The 2013 orbit in heliocentric ecliptic coordinate system

Assuming a timing error of 10 ps and an accelerometer error of $10^{-13}$ m s$^{-2}$ (Hz)$^{-1/2}$ at frequency $f \sim 0.1$ mHz consistent with Fig. 2 and fitting simulated data from 350 days after launch to 750 days after launch, we predict that the uncertainties in determining the relativistic parameters $\gamma$ and $\beta$ to be about $1 \times 10^{-7}$. This simulation supports our original goals.

## 4. Outlook

With the advance of laser technology and the development of space interferometry, one can envisage a 15 W (or more) compact laser power and 2-3 fold increase in pointing ability. With these developments, one can increase the ranging distance from 2 AU for ASTROD to 10 AU (2 × 5 AU) and one can design the spacecraft orbits similar to Jupiter's. Four spacecraft will be ideal for a dedicated gravitational-wave mission (super-ASTROD)[70] to explore the primordial background and lower-frequency gravitational waves.

Space missions using optical devices will be important in testing relativistic gravity and measuring solar-system parameters. Laser Astrodynamics in the solar system envisages ultra-precision tests of relativistic gravity, provision of a new eye to see into the solar interior, precise measurement of $\dot{G}$, monitoring the solar-system mass loss, and detection of low-frequency gravitational waves to probe the early Universe and study strong-field black hole physics together with astrophysics of binaries. One spacecraft and multi-spacecraft mission concepts --- ASTROD I, ASTROD and Super-ASTROD are in line for more thorough mission studies. In view of their importance both in fundamentals and in technology developments, mission concepts along this line will be realized and fruitful.

**Acknowledgements**

I am grateful to all members of the ASTROD/ASTROD I study project for their works and encouragement to make this presentation possible. The ASTROD/ASTROD I study project is supported by voluntary contributions from each participating institute, a 3-year Sino-German ASTROD Study Collaboration Grant from DLR, NAO ASTROD I Study Grant, the Foundation of Minor Planets of Purple Mountain Observatory, and Sino-German Center for Science Promotion. Organizing works from W.-Q. Chao, R. Liu, and Yao Dong to realize this third ASTROD Symposium are highly appreciated. I thank T. Appourchaux, G. Bao, H. Dittus, C. Grimani, G. Heinzel, O. Jennrich, S. Klioner, S. Kopeikin, C. Lämmerzahl, A. Pulido Patón, E. M. Rasel, A. Rüdiger, E. Samain, D. Shaul, S. Shiomi, M. Soffel, S, Theil, L. Wen, A. Wicht, C. Xu, S.-H. Ye, and many others for helpful discussions. I also thank D. Shaul for a critical reading of the manuscript.